\documentclass[aps,superscriptaddress,showpacs,onecolumn]{revtex4}

\usepackage{graphics}
\usepackage{dcolumn}
\usepackage{epsfig}
\usepackage{amsmath}
\usepackage{mathrsfs}
\usepackage{amsfonts}
\usepackage{amssymb}
\begin{document}

\title{\texttt{Phonon Unfolding}: A program for unfolding phonon dispersions of materials}

\author{Fawei Zheng}
\affiliation{Institute of Applied Physics and Computational Mathematics, Beijing, People's Republic of China, zheng\_fawei@iapcm.ac.cn}
\author{Ping Zhang}
\affiliation{Institute of Applied Physics and Computational Mathematics, Beijing, People's Republic of China, zhang\_ping@iapcm.ac.cn}

\date{\today}

\begin{abstract}
We present \texttt{Phonon Unfolding}, a Fortran90 program for unfolding phonon dispersions. It unfolds phonon dispersions by using a generalized projection algorithm, which can be used to any kind of atomic systems in principle. Thus our present program provides a very useful tool for the phonon dispersion and vibration mode analysis of surface reconstructions, atomic point defects, alloys and glasses.
\end{abstract}


\maketitle

{\bf PROGRAM SUMMARY}

\begin{small}
\noindent\\[-0.25cm]
{\em Program Title:} \texttt{Phonon Unfolding}                \\ \\[-0.25cm]
{\em Catalogue identifier:}                                    \\ \\[-0.25cm]
{\em Programming language:} Fortran 90\                        \\ \\[-0.25cm]
{\em Computer:} any computer architecture                      \\ \\[-0.25cm]
{\em RAM: } system dependent, about 10 MB                     \\ \\[-0.25cm]
{\em Program obtainable from: }                               \\ \\[-0.25cm]
{\em Number of processors used:} 1                             \\ \\[-0.25cm]
{\em CPC Library subroutines used:}    None                    \\ \\[-0.25cm]
{\em Operating system:} Linux, Windows, Mac                     \\ \\[-0.25cm]
{\em External routines/libraries:} LAPACK                \\ \\[-0.25cm]
{\em Nature of problem:}
The Brillouin zone of a supercell is smaller than that of a primary cell. It makes the supercell phonons more crowded. The crowded phonon dispersions are outright difficult, if not impossible, to be compared with experimental results directly. Besides, the intra-supercell translation symmetries are hidden in the supercell phonon dispersion calculations. In order to compare with experiments and catch the hidden symmetries, we have to unfold the supercell phonon dispersions into the corresponding primary-cell Brillouin zone.  \\ \\[-0.25cm]
{\em Solution method:}
The phonon polarization vectors are projected to a group of plane waves. The unfolding weight is calculated from these plane wave components.
\\ \\[-0.25cm]
{\em Running time:} system dependent, from a few seconds to one hour \\ \\[-0.25cm]
{\em Unusual features of the program:} Applicable to general systems without considering which kind of translational symmetry breaking. Simple and user-friendly input system. Great efficiency and high unfolding speed. \\ \\[-0.25cm]
{\em References:} \\
 F. Zheng, P. Zhang, \textquotedblleft General methods to unfold phonon dispersions \textquotedblright, arXiv: 1602.06655.

\end{small}

\section{Introduction}
Similar to the electronic Hamiltonian, the dynamical matrix for phonon is invariant under translation $T({\bf n})=n_1{\bf a}_1+n_2{\bf a}\_2+n_3{\bf a}_3$ in an ideal crystal. The ${\bf a}_1\sim {\bf a}_3$ are the lattice vectors of the crystal, and ${\bf n}$ has three integer components ($n_1$,$n_2$,$n_3$). The translational symmetry largely simplifies the crystal physics problems. And the application of Bloch's theorem to atomic vibration problem introduces the concept of phonon dispersions. The phonon dispersions are widely applied in condensed matter physics, especially in exploring the heat transport, atomic structure stabilities and thermal properties in crystal, and is connected to the neutron inelastic scattering experiments.

If the crystal contains translational symmetry breaking, for example the systems with point defects \cite{Heumen, Konbu, Zheng2016}, interfacial reconstruction \cite{Kim, YQi, silicene}, complex spin configurations \cite{KaiLiu}, and disorder \cite{Popescu1,Popescu2,haverkort}, the natural choice in theoretical and computational study is to use the supercell method. However, the widely used supercell calculation scheme folds the phonon dispersions to supercell first Brillouin zone.  The supercell first Brillouin zone is much smaller than that of the primitive cell. All the phonon dispersions become shorter and crowded in the limited reciprocal space. The shapes of phonon dispersions are destroyed, which makes it hard to analyze. The situation would be much more serious for the heavily folded phonon dispersions. The consequent phonon polarization vectors are supercell Bloch functions. Many of them have the same symmetries with primary-cell Bloch functions. It originates from the approximate primary-cell translational symmetry. Furthermore, the neutron inelastic scattering experimental results can not be refereed to the supercell phonon dispersions and catch the hidden translational symmetry. Like the unfolded electron energy bands, we have to unfold the supercell phonon dispersions into the corresponding primitive cell first Brillouin zone.

Recently, the rapidly developed electron energy bands unfolding methods\cite{Boykin1,Boykin2,Allen,Ku,Konbu2011,Popescu1,Popescu2,Lee,huang,Zheng2015,Rubel,Deretzis,Brommer,Farjam} stimulated the study of phonon unfolding methods\cite{Allen,huang,Boykin2014,Zheng2016_2}. In one of our previous papers\cite{Zheng2016_2}, we proposed a generalized calculation scheme to unfold phonon dispersions, which can be applied to atomic systems with any kind of translational symmetry broken. Lets denote the phonon polarization vector as,
\begin{eqnarray}\label{wavefunction}
    \psi^s_{\vec{q}}(\vec{r}_i)&=&\phi^s_{\vec{q}}(\vec{r}_i)e^{\bold{i}\vec{q}\cdot\vec{r}_i}\nonumber\\
          (i=&1,&2, ..., N,\,s=1,2,3).\nonumber
\end{eqnarray}
Here, the band index is omitted. The integer $s=1,2,3$ corresponds to $x$, $y$, and $z$ directions respectively. The $\vec{q}$ is the wave number in the supercell first Brillouin zone, and $\vec{r}_i$ denotes the i-th atom position. The integer $N$ is the supercell total atom number. Suppose the supercell contains $n$ primitive cells. Then, the primitive cell first Brillouin zone contains $n$ Brillouin zones of supercell. The projection operator for each supercell Brillouin zone is noted as $\hat{P}_b\,\, (b=1..n)$. As an approximation, we use the plane wave basis to construct the projection operator as,
\begin{equation}
   \hat{P}_b= \sum_{j,s}|w_{j,s,b}><w_{j,s,b}|\nonumber
\end{equation}
In which,
\begin{equation}
   |w_{j,s,b}>=\frac{e^{\bold{i}(\vec{Q}_b+\vec{G}_j)\cdot\vec{r}_i}}{\sqrt{N}}\delta_{s,s'}.\nonumber
\end{equation}
Where the $\vec{Q}_b$ is the reciprocal lattice point of supercell. The $\vec{G}_j$ is the reciprocal lattice point of primitive cell, i.e. it is the wavenumber of plane wave basis. We symmetrically chose finite number of $\vec{G}_j$ around (0, 0, 0) point. Then the unfolding weight can be written as:
\begin{eqnarray}
|c_{\vec{q}+\vec{Q}_b}|^2 &=&<\phi_{\vec{q}}|\hat{P}_b|\phi_{\vec{q}}>\nonumber\\
&=&\sum_{j,s} |B_{j,s}^{\vec{q}+\vec{Q}_b}|^2\nonumber\\
 B_{j,s}^{\vec{q}+\vec{Q}_b}&=&<w_{j,s,b}|\phi_{\vec{q}}>\nonumber\\
 &=& \sum_i\frac{e^{-\bold{i}(\vec{Q}_b+\vec{G}_j)\cdot\vec{r}_i}}{\sqrt{N}}\phi_{\vec{q}}^{s}(\vec{r}_{i}).\nonumber
\label{method_two}
\end{eqnarray}

The present code \textbf{PhononUnfolding} is based on this plane wave projection scheme. It is applicable to any kind of symmetry breaking systems, even if the system is heavily symmetry broken, when the atom correspondence is not available.   Presently, the code contains interface to Quantum Espresso package.

\section{Brief description of the code}

After decompressing the {\it PhononUnfolding.tar.gz} file, one gets a folder named as {\it PhononUnfolding}. There are {\it Readme} file and four sub-folders. They are {\it src}, {\it system}, {\it doc} and {\it examples}. Sub-folder {\it src} contains two Fortran90 source files, {\it src} contains {\it makefile\_windows} and {\it makefile\_linux} which can be used for compiling the code in Windows and Linux operating systems respectively, {\it doc} folder contains a description file of the present code.

In this code, the control parameters are read from the input.dat file, which will be described in detail in the next section. After we obtain the force constants parameters, there are three steps to get the unfolded phonon dispersions. First of all, we have to produce the q-point list in the first Brillouin zone of supercell. The parameter {\it calculation } is set to {\it qp}. Then the program reads data block {\it begin primary cell qpoint} $\sim$ {\it end primary cell qpoint}. It generates q-point list along the high symmetry lines in the primitive cell Brillouin zone. Then, the program read data block  {\it begin primitive cell vectors} $\sim$ {\it end primitive cell vectors} and {\it begin super cell vectors} $\sim$ {\it end super cell vectors}. From the supercell and primitive cell vectors, the program calculate the q-point correspondence between them, and calculate the q-point list in supercell Brillouin zone. The q-point list is written in file {\it q-list.dat}. If parameter {\it write\_q\_correspondence = true}, then file {\it Q-points.dat} will be produced, which contains more details of the q-point list.

\begin{figure}
\includegraphics[width=0.5\columnwidth]{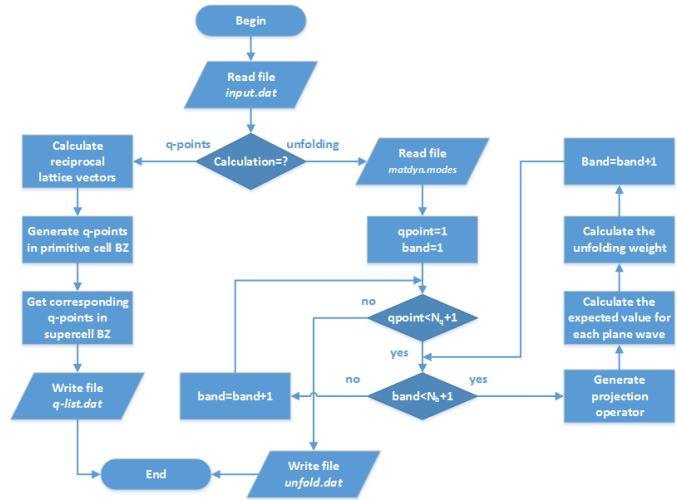}
\caption{\label{Fig1} Flow chart of phonon dispersions unfolding process in \texttt{Phonon Unfolding}. The BZ is the abbreviate of Brillouin zone.}
\end{figure}

Then, we use the q-point list to calculate the supercell phonon polarization vectors. At the last step, we set {\it calculation = uf}. The {\it PhononUnfolding} code reads the polarization vectors produced by $\it matdyn.x$ in Quantum Espresso package. Interfaces to other first-principle and classical molecular dynamics codes will be produced in the near future. The {\it PhononUnfolding} program read parameters {\it max\_qx, max\_qy} and {\it max\_qz}, and gets the plane wave basis. The basis functions are used to get the projection operator $\hat{P}_b$. From which, we finally obtain the phonon unfolding weight. The calculation results are stored in file {\it unfold.dat}, which can be plotted by using Origin or Gnuplot.
The detailed flow chart of unfolding process is shown in Fig. \ref{Fig1}.

\section{The input.dat file}
The target to be performed and the information of the system can be described by the keywords in input.dat file.
The ordering of the keywords is not significant. Case is ignored, so that {\bf calculation} is the same as {\bf Calculation}. Characters after ! or \# are treated as comments. Most keywords have default values unless they are given in input.dat.
The keywords are described as follows:

\begin{itemize}
\item{ {\bf calculation} = uf $\left.\right|$ qp}
    \\ {\it Default value} : uf
    \\ {\it Value type} : string of characters
    \\[-0.25cm]
    \\The keyword {\bf `calculation'} describes the task to be performed. The value of {\bf `calculation'} has two options at the present time; they are:
    \\[-0.25cm]
    \\uf :  phonon dispersion unfolding.
    \\[-0.25cm]
    \\qp : q-point list generation.

\item{{\bf max\_qx},{\bf max\_qy},{\bf max\_qz} = 0, 1, 2 ...}
\\{\it Default value} : 2
\\ {\it Value type} : integer number
\\[-0.25cm]
\\ {\bf max\_qx}, {\bf max\_qy}, and {\bf max\_qz}  define the plane wave groups to be used in projection calculation.  The wave numbers are: -{\bf max\_qx}$<n_x<${\bf max\_qx}, -{\bf max\_qy}$<n_y<${\bf max\_qy} and -{\bf max\_qz}$<n_z<${\bf max\_qz}. For low dimensional systems, the non-periodic dimension value can be simple set to zero. For example,  {\bf max\_qz} = 0 in graphene system in x-y plane.

\item{{\bf wtclean} = 0 $\sim$ 1}
\\{\it Default value} : 0.01
\\ {\it Value type} : real number
\\[-0.25cm]
\\The result data point with unfolding weight lower than {\bf `wtclean'} will not be written to the output file. Then the output file would be smaller and is easier to handle.

\item{{\bf write\_q\_correspondence}= t $\left.\right|$ true $\left.\right|$ f $\left.\right|$ false}
\\{\it Default value} : false
\\ {\it Value type} : logic
\\[-0.25cm]
\\ The value is case insensitive. If {\bf `write\_q\_correspondence'} = true or t, then a {\it Q-points.dat} file is produced, which contains all the q point list in primitive cell Brillouin zone and the corresponding q point list in super cell Brillouin zone. There are seven columns in {\it Q-points.dat} file. The first column is the high symmetry path length, the next three columns are q point list in first Brillouin zone of primitive cell, the last three columns are the corresponding q point list in the first Brillouin zone of supercell.

\item{{\bf begin primary cell qpoint} and {\bf end primary cell qpoint}}
\\[-0.25cm]
\\ The data block between {\bf `begin primary cell qpoint'} and {\bf `end primary cell qpoint'} defines the high-symmetry lines in the first Brillouin zone of the primitive cell. There are $3n$ lines in the data block. Each three lines define one high-symmetry q-point line. The first line is an integer, which is larger than 1. It shows the number of q-points along the high-symmetry q-point line. Both the second and the third lines have three real numbers, which show the starting and end points of the high-symmetry q-points line in direct form.

\item{{\bf begin primitive cell vectors} and {\bf end primitive cell vectors}}
\\[-0.25cm]
\\ The data block between {\bf `begin primary cell vectors'} and {\bf `end primitive cell vectors'} have three lines. They define the three cell vectors of primitive cell with the unit of angstrom.

\item{{\bf begin super cell vectors} and {\bf end super cell vectors}}
\\[-0.25cm]
\\ The data block between {\bf `begin super cell vectors'} and {\bf `end super cell vectors'} have three lines. They define the three cell vectors of super cell with the unit of angstrom.

\item{{\bf begin super cell atom positions} and {\bf end super cell atom positions}}
\\[-0.25cm]
\\ The data block between {\bf `begin super cell atom positions'} and {\bf `end super cell atom positions'} show the atom positions of supercell with the unit of angstrom. Each line have three real numbers which is the Cartesian position of the corresponding atom. The order of the atoms in this data block should be the same with that of phonon polarization vector file {\it matdyn.modes}.

\end{itemize}

\section{Examples}
In the following context we illustrate the capabilities of \texttt{Phonon Unfolding} by describing two systems: (i) graphene, a two dimensional system without translational symmetry breaking; (ii) diamond with a carbon vacancy, a three dimensional system with translational symmetry breaking;
The unfolding process is trivial in graphene. The calculation results should coincide with the phonon dispersions from the primitive-cell calculations. The example here shows the validity of the present code. The translational symmetries in diamond with carbon vacancy are broken. Then, the unfolding process is nontrivial.

\subsection{Graphene}
Since the success of fabrication single layer graphene\cite{Novoselov2004}, this material has been drawing much attentions\cite{yan,Mohanty,Liao,Xia,graphene2015,graphene2016} due to its peculiar properties\cite{graphene_rmp,Geim,graphene.fab,dirac.nature,dirac.nature2}. Similar to graphite, {\it h}-BN, and carbon nanotubes, graphene has two-dimensional honeycomb structure. It is the building block of fullerene, carbon nanotubes, graphene nanoribbons, and graphite.

\begin{figure}
  \includegraphics[width=0.5\columnwidth]{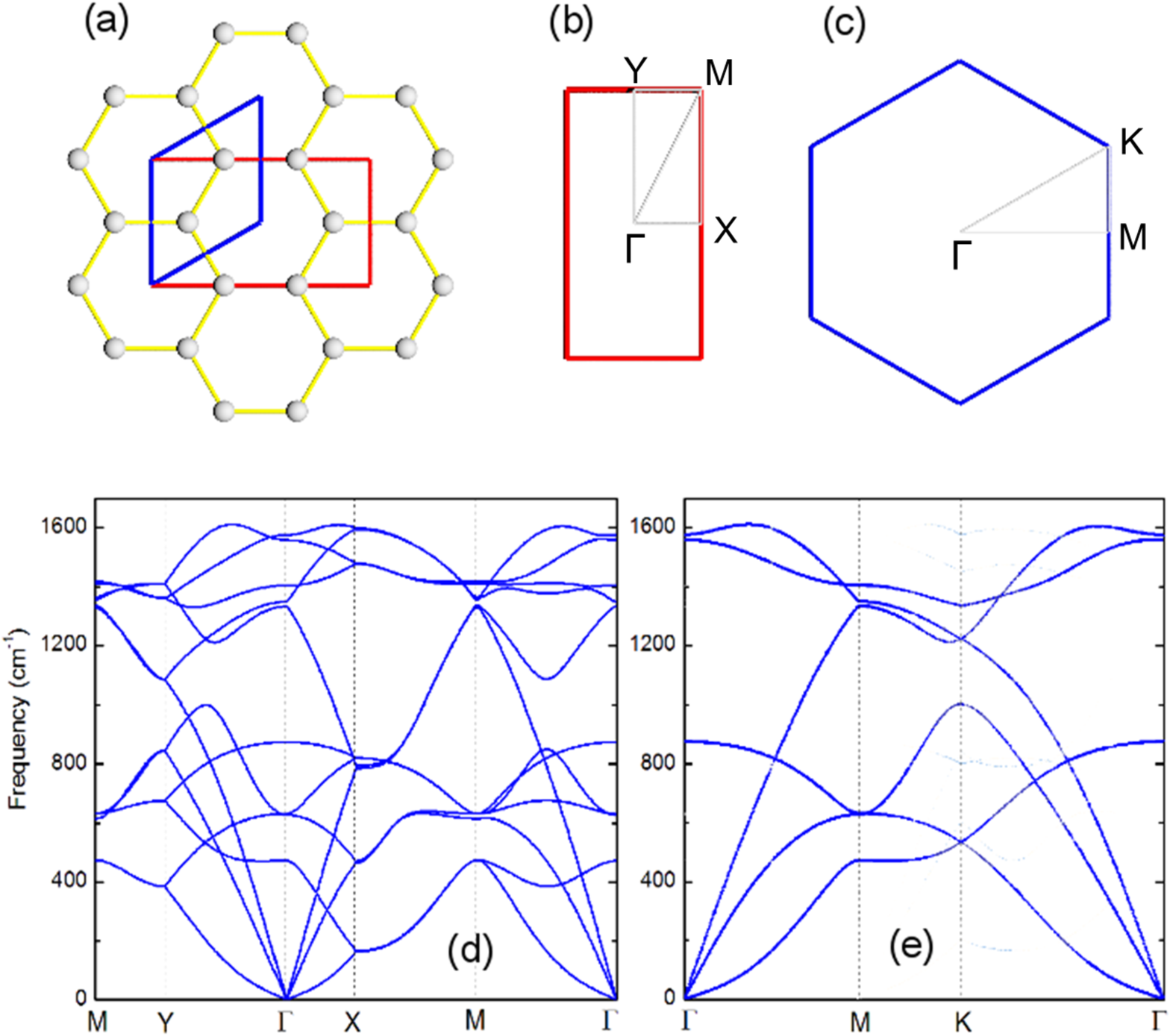}
  \caption{
  \label{Fig2}
  (a) Atomic structure of graphene with its supercell (red rectangle) and primary cell (blue parallelogram). Panels (b) and (c) show the first Brillouin zones of the supercell and the primary cell, respectively, with high symmetry points and lines. Panel (d) shows the phonon dispersions of the supercell. Panel (e) shows the phonon dispersions unfolded into the primary-cell Brillouin zone. }
\end{figure}

As an example, we consider phonon dispersion unfolding of a freestanding graphene. The structure relaxation and electronic structure calculations are performed by using DFT \cite{DFT1,DFT2} with  norm-conserving carbon pseudopotential \cite{NC1,NC2}. The exchange correlation potential is described by the generalized gradient approximation (GGA) of Perdew-Burke-Ernzerhof (PBE) type \cite{PRL.77.3865}. The kinetic energy cutoff for wavefunction is chosen to be 120 Ry, which is converged in our test. We use a rectangle supercell in our DFT calculations. Each supercell contains four carbon atoms as shown in Fig. \ref{Fig2}. The graphene layers are separated by a vacuum of 20 \AA\, in order to reduce the interactions between the nearest layers. The system is relaxed until the force on each atom is smaller than 0.005 eV/\AA. BFGS quasi-newton algorithm is used in the structure relaxation. In the self-consistent ground state calculations, 9$\times$16$\times$1 Monkhorst-Pack K-points setting is used in the reciprocal space integration. After obtaining the self-consistent ground state, we performed the density functional perturbation theory (DFPT) calculations with a uniform 3$\times$5$\times$1 grid of q-points setting. Then perform dynamical matrix Fourier transformations to get force constants, and calculated the phonon dispersions in graphene supercell. All the DFT and DFPT calculations are performed by using Quantum Espresso package \cite{pwscf}.

The atomic structure of graphene is shown in Fig. \ref{Fig2}(a). In which, the red rectangle and blue parallelogram are the supercell and primitive cell respectively. The supercell and primitive cell contain 4 and 2 carbon atoms respectively. Their first Brillouin zone and the high symmetry lines used in calculations are shown in Fig. \ref{Fig2}(b,c). The calculated phonon dispersions in graphene supercell are shown in Fig. \ref{Fig2}(d).
Then we calculate the q points list in primitive cell Brillouin zone high-symmetry lines, and the corresponding q points list in supercell Brillouin zone. This can be down by setting {\it calculation = qp } in {\it input.dat} file. After performing {\it PhononUnfolding.exe}, the q points list is stored in the {\it q-list.dat} file. We copy the list to the input file of {\it matdyn.x } in Quantum Espresso package. And calculate the phonon dispersions and polarization vectors (stored in {\it matdyn.modes} file) in supercell. After that, we set {\it calculation = uf } in {\it input.dat}, and run {\it PhononUnfolding.exe} again. The calculation results are written in {\it unfold.dat} file. The plotted unfolded phonon dispersions are shown in Fig. \ref{Fig2}(e), which is the same with phonon dispersions in primitive cell graphene\cite{saito}.

\subsection{Diamond with Vacancy}
\begin{figure}
  \includegraphics[width=0.5\columnwidth]{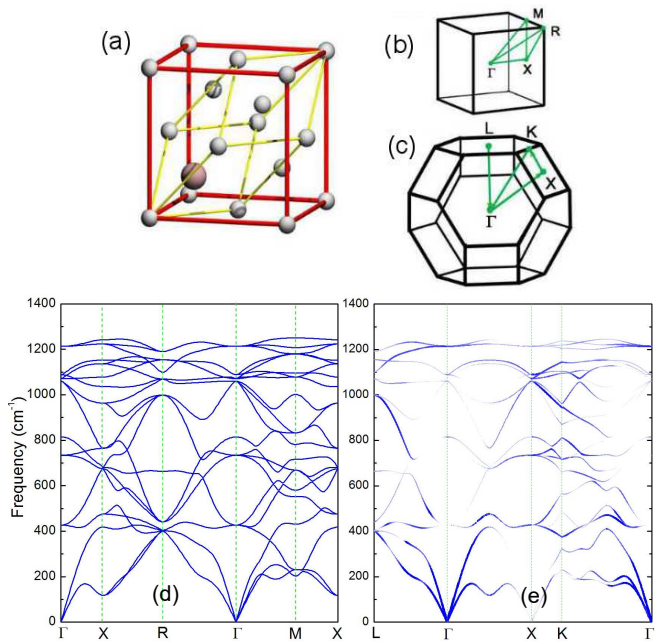}
  \caption{
  \label{Fig3}
   (a) Atomic structure of diamond with one vacancy. The supercell and primitive cell are shown by a red and yellow cages. The C atoms and vacancy are shown by gray and brown balls, respectively. Panels (b) and (c) show the first Brillouin zones of supercell and  primitive cell with high symmetry lines. Panels (d) shows the phonon dispersions of the supercell, and panel (e) shows the phonon dispersions unfolded into the primitive cell Brillouin zone.
  }
  \end{figure}

In order to test the code in three-dimensional systems, we further calculate the phonon dispersions of a diamond supercell with a carbon vacancy, and unfold it to the primary-cell Brillouin zone. The DFT calculations are carried out to relax the atomic structure. The inner electrons of carbon atoms are described by norm-conserving pseudopotentials \cite{NC1,NC2}. The exchange correlation potential is described by the GGA of PBE-type \cite{PRL.77.3865}. The kinetic energy cutoff for wavefunction is chosen to be 120 Ry, which is converged in our test. We use a cubic supercell in our DFT calculations as shown in Fig. \ref{Fig3}(a). Each supercell contains seven carbon atoms and one carbon vacancy. The carbon atoms and vacancy are shown by gray and brown balls respectively. The primary cell is also shown in Fig. \ref{Fig3}(a) by a yellow cage. The first Brillouin zone and high symmetry lines used in our calculations of supercell and primitive cell are shown in Fig. \ref{Fig3}(b) and (c) respectively. The relaxed supercell lattice parameter is 3.57 \AA. The system is relaxed until the force on each atom is smaller than 0.005 eV/\AA. BFGS quasi-newton algorithm is used in the structure relaxation. In the self-consistent ground state calculation, a 13$\times$13$\times$13 Monkhorst-Pack K-points setting is used in the reciprocal space integration. After obtaining the self-consistent ground state, we performed the DFPT calculations with 4$\times$4$\times$4 grid of q-point setting. The force constants were obtained by Fourier transport of dynamic matrix. The supercell phonon dispersions calculated by using the force constants are shown in Fig. \ref{Fig3}(d). They are complex compare to the primitive cell phonon dispersions due to the additional atoms and an extra vacancy.

Then we continue to calculate the unfolded phonon dispersions. Similar to the case of graphene, we first generate the q-point list by setting {\it calculation = qp } in {\it input.dat} file. After running {\it PhononUnfolding.exe}, the q points list is stored in the {\it q-list.dat} file. We copy the q-list to the input file of {\it matdyn.x }. Then we calculate the phonon dispersions and polarization vectors in supercell. The phonon polarization vectors are stored in {\it matdyn.modes} file. After that, we set {\it calculation = uf } in {\it input.dat}, and run {\it PhononUnfolding.exe} again. The program loads the phonon polarization vectors and produces the phonon dispersion unfolding weight. The calculation results are written in {\it unfold.dat} file. The plotted unfolded phonon dispersions are shown in Fig. \ref{Fig3}(e). The unfolded energy bands of the doped system have broken points and darkness in a variety, which originate from the breaking of translational symmetry.

\section{Conclusion}

In this communication we introduced \texttt{Phonon Unfolding}, a computer code for
unfolding phonon dispersions by using. \texttt{Phonon Unfolding} enables accurate and efficient
calculations of the phonon dispersions. The executable versions of Phonon Unfolding for Windows and Linux operation systems are distributed by email.  Plans are in place to extend \texttt{Phonon Unfolding} in order to unfold phonon dispersions calculated by Phon\cite{phon}, Phonon\cite{phonon}, Phonopy\cite{phonopy}, ABINIT\cite{abinit}, Siesta\cite{siesta}, and LAMMPS\cite{lammps}.

\section{Acknowledgments}

The research leading to these results has received funding from Natural Science Foundation of China under Grants No.11474030 and the joint grant of NSFC and NSCC-GZ.

\end{document}